\documentclass{article}

\usepackage{arxiv}

\usepackage[utf8]{inputenc} 
\usepackage[T1]{fontenc}    
\usepackage{hyperref}       
\usepackage{url}            
\usepackage{booktabs}       
\usepackage{amsfonts}       
\usepackage{nicefrac}       
\usepackage{microtype}      
\usepackage{lipsum}
\usepackage{graphicx}
\graphicspath{./Figure/}
\usepackage{subcaption}
\usepackage{ragged2e}
\usepackage{tabularx}
\usepackage{amsmath}
\usepackage{mathtools}
\usepackage{cuted}
\usepackage{amssymb}
\usepackage{lipsum}
\usepackage{nccmath}
\usepackage{autobreak}
\usepackage{makecell}
\usepackage{xurl}
\usepackage{xcolor}

 \title{Unveiling the Inter-Related Preferences of Crowdworkers: Implications for Personalized and Flexible Platform Design}

\author{
 Senjuti Dutta \\
  University of Tennessee, Knoxville\\
  \texttt{sdutta6@vols.utk.edu} \\
   \And
 Rhema Linder \\
University of Tennessee, Knoxville\\ 
  \texttt{rlinder@utk.edu} \\
  \And
 Alex C. Williams \\
  Amazon AI \\
  \texttt{acwio@amazon.com} \\
 \And
Anastasia Kuzminykh \\
    University of Toronto\\
    \texttt{anastasia.kuzminykh@utoronto.ca} \\
    \And
Scott Ruoti \\
   University of Tennessee, Knoxville\\ 
     \texttt{ruoti@utk.edu}  \\
}
\begin{document}

\maketitle

\begin{abstract}

Crowdsourcing platforms have traditionally been designed with a focus on workstation interfaces, restricting the flexibility that crowdworkers need. Recognizing this limitation and the need for more adaptable platforms, prior research has highlighted the diverse work processes of crowdworkers, influenced by factors such as device type and work stage. However, these variables have largely been studied in isolation. Our study is the first to explore the interconnected variabilities among these factors within the crowdwork community.
Through a survey involving 150 Amazon Mechanical Turk crowdworkers, we uncovered three distinct groups characterized by their interrelated variabilities in key work aspects. The largest group exhibits a reliance on traditional devices, showing limited interest in integrating smartphones and tablets into their work routines. The second-largest group also primarily uses traditional devices but expresses a desire for supportive tools and scripts that enhance productivity across all devices, particularly smartphones and tablets. The smallest group actively uses and strongly prefers non-workstation devices, especially smartphones and tablets, for their crowdworking activities.
We translate our findings into design insights for platform developers, discussing the implications for creating more personalized, flexible, and efficient crowdsourcing environments. 
Additionally, we highlight the unique work practices of these crowdworker clusters, offering a contrast to those of more traditional and established worker groups.
\end{abstract}
\section{Introduction}

One of the primary reasons crowdworkers are drawn to crowdwork is the flexibility it offers. However, despite this critical advantage, the current architecture of crowdwork platforms significantly limits the flexibility and autonomy workers possess, particularly in orchestrating their work schedules and selecting their preferred work environments~\cite{newlands2021crowdwork,dutta2022mobilizing}. This rigidity starkly contrasts with the inherently fluid and varied nature of crowdwork, highlighting an urgent need for these platforms to evolve to better support the diverse and dynamic work practices of crowdworkers.

To address this need, it's essential to dig into the core characteristics of work processes that vary among crowdworkers. A standard workflow for crowdworkers includes stages such as managing tasks, which involves finding and accepting work, as well as completing various types of tasks like information retrieval and classification~\cite{williams2019perpetual,newlands2021crowdwork,gadiraju2014taxonomy}. Previous research has identified device type~\cite{dutta2022beyond} and work stage, including task completion and management~\cite{williams2019perpetual}, as critical aspects influencing work practices. While prior studies have illuminated crowdworkers' preferences for utilizing a range of devices beyond traditional computers for both completing and managing tasks~\cite{hettiachchi2020context,williams2019perpetual}, a significant gap remains in understanding whether there are systematic preferences among crowdworkers for choosing specific device types at different work stages.

In this paper, we aim to bridge this knowledge gap by conducting a survey of 150 Amazon Mechanical Turk (MTurk) workers. We investigate the current practices and desired preferences of crowdworkers regarding the use of various devices. Building on previous research, we focus on four key types of non-workstation devices, in addition to workstations, that have garnered significant interest: smartphones~\cite{dutta2022mobilizing,vashistha2017respeak,chi2018mobile}, tablets (including situated tablets)\cite{hosio2014situated,goncalves2015motivating,goncalves2017task}, smart speakers\cite{hettiachchi2019enabling,hettiachchi2020hi}, and smartwatches~\cite{nebeling2016wearwrite,acer2019scaling}, for both completing and managing tasks. Through this exploration, we uncover three distinct clusters of workers providing a deeper understanding of the diverse ecosystem of digital labor.

Our contributions include identifying three distinct groups of crowdworkers with varying work practices based on device usage and work stages:
\begin{itemize}
	\item  \textbf{Cluster 1 (n = 66)}: The largest group, consisting of workers who do not desire to use non-workstation devices for task completion and are highly skeptical about supporting management on all non-workstation devices. This indicates their dependence on traditional devices, though they exhibit limited openness to using smartphones and tablets for completing and managing tasks.
	
	\item \textbf{Cluster 2 (n = 47)}: The second-largest group do not favor using non-workstation devices for task completion in the future, indicating a reliance on traditional devices. Desite this, they have openness using non-workstation devices currently as well as 
	they want support for scripts and tools to manage tasks across all devices, especially for smartphones and tablets.
	
	\item \textbf{Cluster 3 (n = 35)}: The smallest group consists of non-workstation enthusiasts who prefer using non-workstation devices, particularly smartphones and tablets, for both managing and completing tasks. Additionally, they express a strong desire for enhanced management support across all devices.

\end{itemize}

Through these insights, we offer design recommendations for creating more personalized, flexible, and efficient crowdsourcing platforms, fostering a more adaptable and worker-friendly environment. Additionally, we compare the unique work practices of these clusters of crowdworkers with more traditional and established worker groups to better understand their work practices.

\section{Background}
This section provides the background information necessary to supplement the understanding of the work presented in this paper. We offer a comprehensive review of studies examining how different device types have been used by crowdworkers across various stages of their tasks.

\subsection{Use of Workstation in Different Stages of Crowdwork}
Crowdwork, a work practice centered around the completion of tasks for payment, primarily relies on workstation computers, as evidenced by numerous studies. Both quantitative and qualitative research consistently shows that the vast majority of crowdworkers, similar to other information work professions, identify workstation and laptop computers as their primary devices for work-related activities~\cite{hettiachchi2020context,williams2019perpetual}.

In addition to task completion, crowdworkers engage in various activities to manage their work, such as finding HITs, communicating with other crowdworkers, and reviewing requesters~\cite{martin2014being,williams2019perpetual,gray2019ghost,savage2020becoming,toxtli2020reputation}. Much of this work is driven by voluntary or community-based efforts where crowdworkers self-manage infrastructure to enhance their productivity. Examples include platforms for reviewing HITs or requesters like Turkopticon~\cite{irani2013turkopticon}, forums for connecting with other crowdworkers like TurkerView~\cite{savage2020becoming}, and the development and sharing of new productivity tools~\cite{williams2019perpetual}. Typically, these managmeent tools are designed with workstation computers as the intended use case.

\subsection{Use of Non-workstation Devices in Different Stages of Crowdwork}
Prior research has also highlighted the potential of non-workstation devices in crowdsourcing tasks. Early examples include TxtEagle~\cite{eagle2009txteagle} and mClerk~\cite{gupta2012mclerk}, which allowed individuals in Kenya and India to complete tasks via SMS, such as language translation, market research, and audio transcription, using basic mobile phones. Building on this, Narula et al. developed MobileWorks, a smartphone platform for optical character recognition~\cite{narula2011mobileworks}, showcasing the capability of smartphones for completing more complex and varied crowdwork tasks.

Voice-based interactions have further expanded the use of mobile devices in crowdwork. For instance, Vashistha et al.~ developed the Respeak system for assisted transcription~\cite{vashistha2017respeak} , while VoiceTranscriber~\cite{lee2015voicetranscriber} is a mobile crowd-powered system that summarizes stories from recorded voices, leveraging human abilities of discrimination and expression. 
Beyond voice-based interactions, Google’s Crowdsource app enables image transcription, translation, and handwriting recognition tasks via smartphones~\cite{chi2018mobile}. Additionally, Chopra et al. created an Android app for digitizing handwritten Marathi/Hindi words, emphasizing the capability of mobile devices for specific linguistic tasks~\cite{chopra2019exploring}. Research by Williams et al. and Newlands et al. investigated smartphone use for crowdwork management, such as finding tasks, creating catchers, and monitoring tasks\cite{williams2019perpetual,newlands2021crowdwork}.

Research has explored the multi-device landscape of crowdwork beyond smartphones. Previous research demonstrated that tablets could be used for tasks in public spaces~\cite{goncalves2013crowdsourcing,goncalves2016crowdsourcing} . Researchers also explored the use of smart speakers for HIT completion, finding them effective for multitasking~\cite{hettiachchi2019enabling,hettiachchi2020hi}. 
In the realm of wearable technology, this paper introduced a smartwatch app to improve task response rates and accuracy for mobile postal workers~\cite{acer2019scaling}. Furthermore, previous work showed how smartwatches could be used to delegate and manage complex tasks, formalizing management patterns in crowdwork~\cite{calacci2022} .

\subsection{Relation to This Work}
Previous research demonstrates that workstations, smartphones, tablets, smart speakers, and smartwatches can all be utilized to manage and complete crowdwork. However, prior studies do not address the systematic differences between workers based on their device choices at each stage of work. This paper aims to fill this gap by exploring whether there are identifiable differences among crowdworkers in their device preferences throughout the various stages of their tasks.

\section{Methodology}


\begin{figure}
	\centering    
      \includegraphics[width=0.8\linewidth]{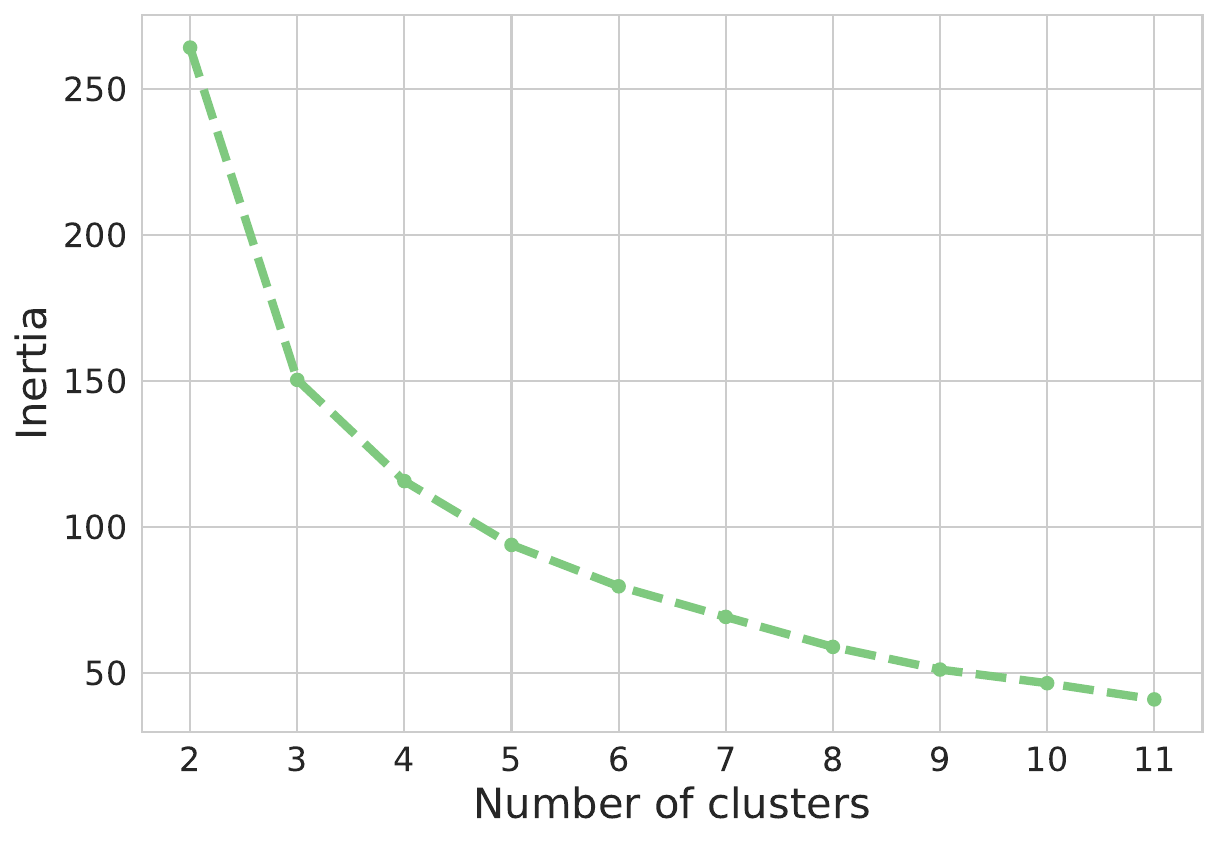}
 
	\caption{Elbow plot which shows the optimized number of clusters are 3}
	\label{fig:elbow_plot}
\end{figure}
\begin{figure}[t]
	\centering
    \includegraphics[width=\linewidth]{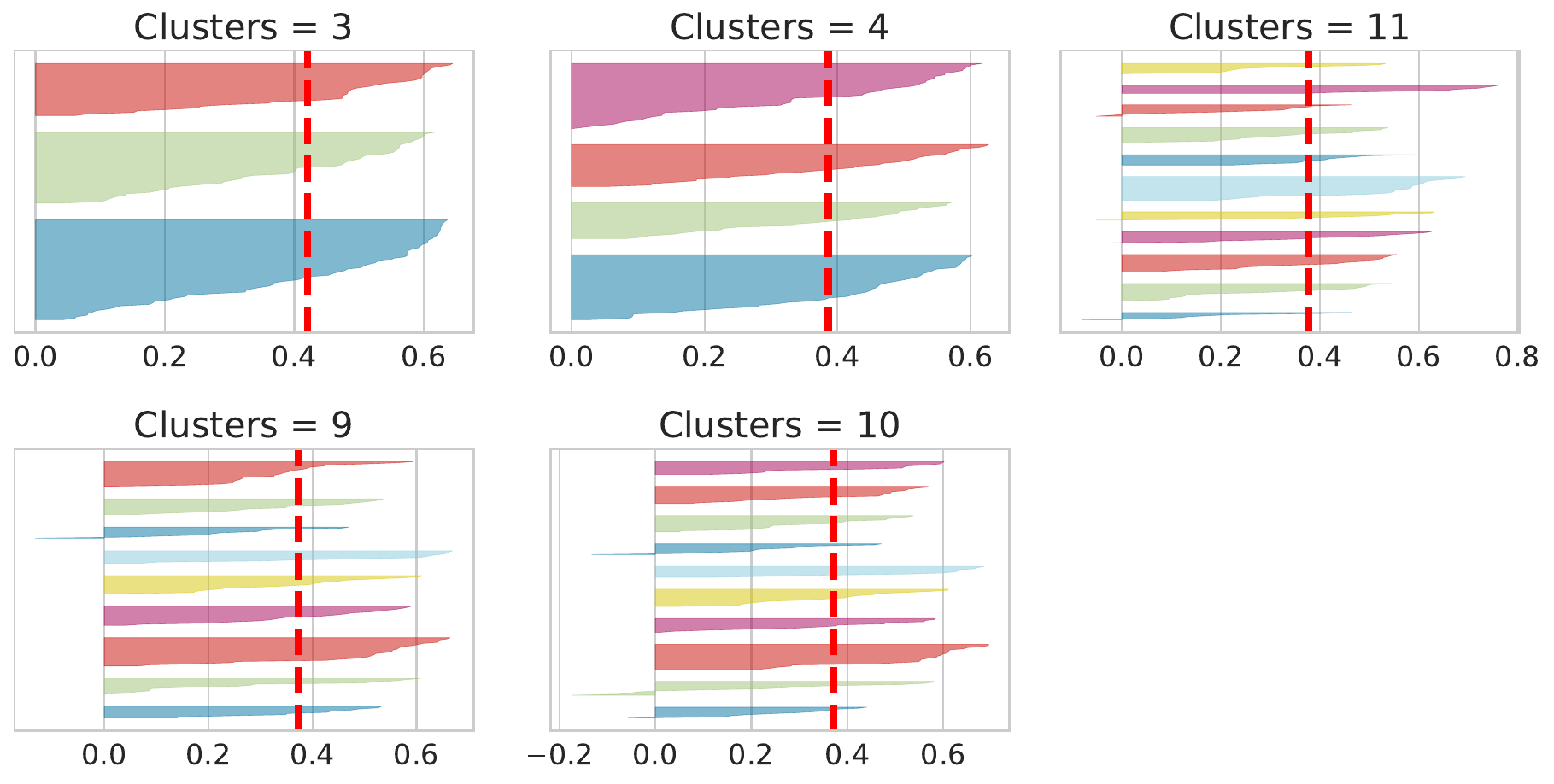}
 
	\caption{Siloutte score plot which shows the optimized number of clusters at 3 }
	\label{fig:silhouette_plot}
\end{figure}
\begin{figure}
	\centering
     \includegraphics[width=\linewidth]{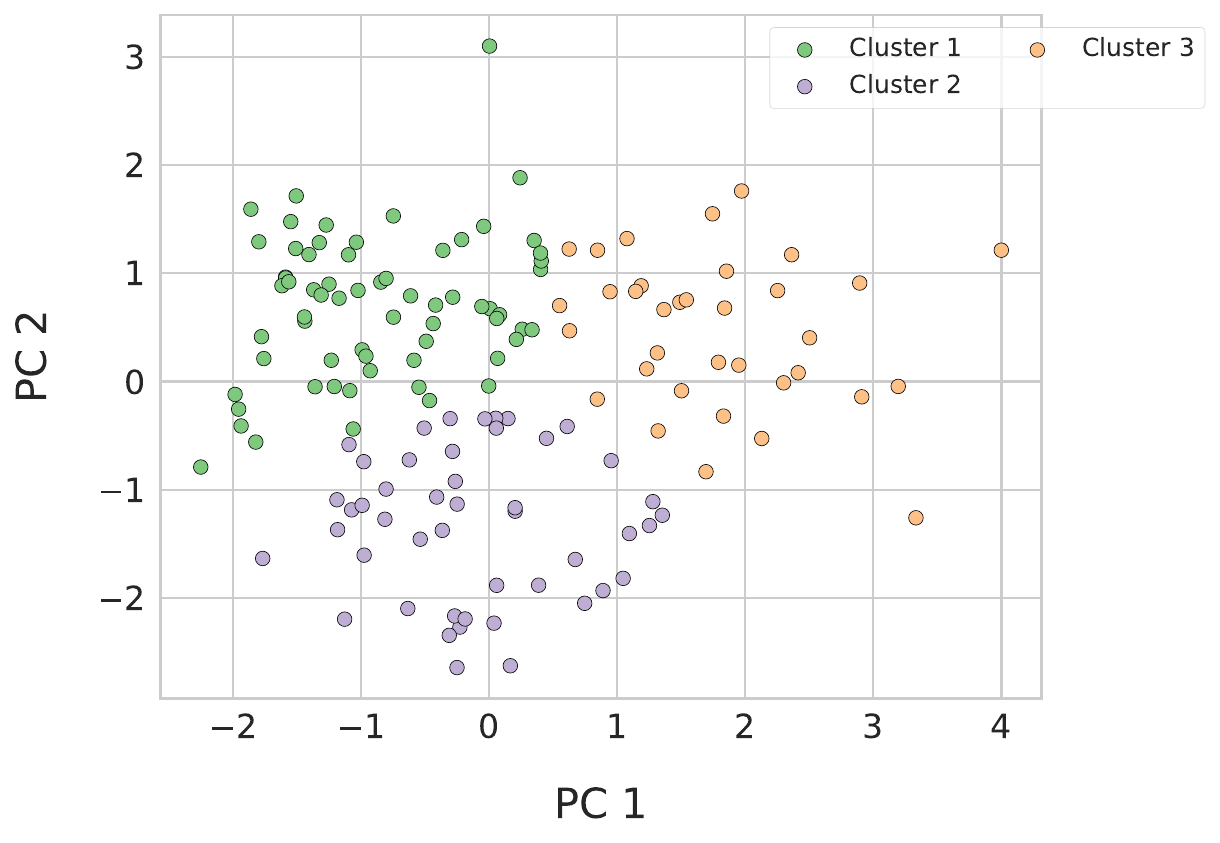}
	\caption{Visualization of crowdworker clusters with the PCA component}
	\label{fig:pca_clusters}
\end{figure}

\begin{table}
\centering
    \begingroup 
        \fontsize{9}{10}\selectfont 
		\begin{tabular}{l| l}
			\hline
			\textbf{Feature}                                               & \textbf{LF} \\ \hline
			Desired\_support\_Management\_Tablet & 0.21 \\  \hline
			Desired\_support\_Management\_Phone  & 0.20 \\ \hline
			Desired\_support\_Management\_Watch   & 0.19 \\ \hline
			Desired\_support\_Management\_Speaker   & 0.19 \\ \hline
			Magic\_wand\_Tablet\_Management\_script & 0.18 \\ \hline
			Magic\_wand\_HIT\_Management\_script\_any\_device & 0.18 \\ \hline
			Desired\_HIT\_type\_Tablet\_wont\_use & -0.15 \\ \hline
			Desired\_HIT\_type\_Watch\_wont\_use & -0.15 \\ \hline
			Desired\_HIT\_type\_Speaker\_wont\_use & -0.15 \\ \hline
			Desired\_HIT\_type\_Phone\_wont\_use & -0.13 \\ \hline
			Magic\_wand\_Phone\_HIT\_Management\_script & 0.13 \\ \hline
			Current\_HIT\_type\_Tablet\_dont\_use & -0.12 \\ \hline
			Magic\_wand\_HIT\_UX\_support\_any\_device & 0.11 \\ \hline
			Desired\_HIT\_type\_Phone\_Survey & 0.10 \\ \hline
			Desired\_HIT\_type\_Tablet\_Survey & 0.10 \\ \hline
			Magic\_wand\_HIT\_Management\_filtering\_any\_device & 0.10 \\ \hline
	\end{tabular}
    \endgroup

    \caption{Top features of PCA Component 1 with loading factors within absolute value of 0.10}
	\label{tab:pca_component_1_features}
\end{table}

\begin{table}	
    \centering
    \begingroup 
        \fontsize{9}{10}\selectfont 

        \begin{tabular}{l | l}
            \hline
            \textbf{Features }                                               & \textbf{LF} \\ \hline
            Rate\_Tablet\_Current\_Management                                     & 0.20           \\ \hline
            Rate\_Phone\_Current\_Management                                      & 0.18           \\ \hline
            Magic\_wand\_HIT\_Management\_script\_Phone  & -0.18          \\ \hline
            Rate\_Tablet\_Current\_Completion                                     & 0.17           \\ \hline
            Current\_HIT\_type\_Watch\_dont\_use                     & 0.15           \\ \hline
            Current\_HIT\_type\_Tablet\_dont\_use                    & 0.14           \\ \hline
            Rate\_Phone\_Current\_Completion                                      & 0.13           \\ \hline
            Current\_HIT\_type\_Phone\_Device\_specific\_only        & -0.13          \\ \hline
            Magic\_wand\_HIT\_Management\_script\_any\_device    & -0.12          \\ \hline
            Magic\_wand\_HIT\_Management\_script\_Tablet & -0.12          \\ \hline
            Current\_HIT\_type\_Speaker\_dont\_use                   & 0.11           \\ \hline
            Desired\_Support\_Management\_Tablet                           & -0.11          \\ \hline
            Rate\_Watch\_Current\_Management                                      & 0.10           \\ \hline
        \end{tabular}
    \endgroup

    \caption{Top features of PCA Component 2 with loading factors within absolute value of 0.10}
	\label{tab:pca_component_2_features}
\end{table}


To understand the differences in crowdworkers' work practices, we conducted a survey of 150 MTurk workers. The survey gathered participants' demographics and inquired about their current and desired practices for completing and managing tasks using workstations, smartphones, tablets, speakers, and watches on crowdsourcing platform (See survey details\footnote{\url{https://github.com/sduyr/Survey/tree/main}}). After excluding two participants who provided incoherent responses, we analyzed data from 148 participants.

\subsection{Analysis  Method}
Qualitative responses were analyzed using thematic analysis~\cite{boyatzis1998transforming}. 
Categorical responses were first converted into discrete numerical values. Subsequently, each qualitative response was encoded into binary values to represent the presence or absence of each identified code. 
We then used Principal Component Analysis (PCA) to reduce the dataset's dimensionality while preserving its core structure and patterns. This allowed us to analyze crowdworkers' practices more effectively. The data were normalized for PCA, and we selected three components using the Kaiser rule, focusing on those with eigenvalues above 1.0. 
Features for each component were chosen based on a threshold of 0.10 for loading scores.
This threshold, representing 10\% of the loading factor, was determined to balance significance and noise reduction, ensuring robust and meaningful feature selection.

Following PCA, we applied k-means clustering (see Figure ~\ref{fig:pca_clusters} for the clusters visualized with two PCA components). 
We clustered in a lower-dimensional space using PCA to reduce noise and enhance cluster separation. 
However, we used the entire original dataset to describe the clusters for statistical differences and decision tree which is discussed in detail below.
The elbow method suggested three clusters as the optimal number (see Figure~\ref{fig:elbow_plot}), which was further supported by a silhouette score plot indicating strong cluster cohesion and separation (see Figure~\ref{fig:silhouette_plot}). Thus, we identified three distinct clusters of crowdworkers based on their work practices.
Subsequently, we describe each cluster using a specialized approach.
%
%
%
%

\begin{table}[t]
    \centering
    \begingroup 
        \fontsize{9}{10}\selectfont 
        \begin{tabular}{l |l |l |l |l}
            \textbf{\makecell{PCA\\Comp. 1}} & \textbf{\makecell{PCA\\Comp. 2}} & \textbf{Cluster} & \textbf{\makecell{Size}} & \textbf{Statistics} \\ \hline
            -0.86 & 0.70 & Cluster 1 & 66 & Mean \\ 
            -0.12 & -1.32 & Cluster 2 & 47 &  \\ 
            1.79 & 0.46 & Cluster 3 & 35 &  \\ \hline
            -1.01 & 0.74 & Cluster 1 & 66 & Median \\ 
            -0.25 & -1.23 & Cluster 2 & 47 &  \\ 
            1.74 & 0.66 & Cluster 3 & 35 &  \\ \hline
        \end{tabular}
    \endgroup
	
    \caption{Each PCA component's mean and median values for each cluster along with their sizes}
    \label{tab:pca_component_mean_median_along_with_cluster}
\end{table}

\subsubsection{Describing Clusters Overall Approach}

Explainability in machine learning and artificial intelligence remains a challenging problem. Each method of explaining model behavior has its strengths and weaknesses~\cite{vilone2021notions}. To address this issue comprehensively, we employed a multi-faceted approach, triangulating our explanation using three distinct methods: (1) Principal Component (PCA) Based Analysis, (2) statistical differences, and (3) decision tree. This combined approach allowed us to capture a more holistic and nuanced understanding of the crowdworker clusters.

\paragraph{Principal Component (PCA) Based Analysis}
We describe each cluster using the mean and median values of the PCA component values. PCA helped reduce the dimensionality of our dataset, highlighting the most significant features contributing to the variance within the data. By analyzing the central tendency measures of these components, we characterized each cluster effectively, providing a foundation for further statistical analysis.
\paragraph{Statistical Differences}
Next, we examined the presence of significant differences among the clusters for each feature using appropriate statistical tests based on the data type. For categorical features, we applied the chi-square test to determine statistically significant differences. To ensure the robustness of our results, we adjusted the threshold values for significance using the Bonferroni correction method, which accounts for multiple comparisons. Features with statistically significant differences were further analyzed based on their effect sizes, employing a threshold of more than 0.30 to denote medium to large effect sizes~\cite{cohen2013statistical}. This method allowed us to quantify the magnitude of differences between clusters, facilitating a deeper understanding of their unique characteristics.
\paragraph{Decision Tree}
Finally, we employed decision tree analysis to understand the features that differentiate each cluster of crowdworkers. Decision trees provided a clear and interpretable model of how different features contributed to cluster membership. This step enabled us to identify the most critical features that distinguished each cluster, complementing the insights gained from the PCA and statistical tests. 
By integrating these three methods, we triangulated our explanation, providing a comprehensive and nuanced understanding of the crowdworker clusters.

Before applying PCA, we grouped tasks into "completion tasks" and "management tasks" to reduce dimensionality and improve interpretability. For decision tree analysis, we retained specific task types to leverage granularity and accurately assess feature importance, enabling precise classification and deeper insights.

\section{Findings}


Our participants have substantial MTurk experience. The majority were male (M=96; F=52). Over half (n = 75, 51\%) held at least a Bachelor's degree. Most (n = 117, 79\%) had 2+ years of MTurk experience. 38\% (n = 56) worked 10-20 hours per week, and around 21\% (n = 31) worked 30+ hours. The median number of HITs was 24,170, with a median approval rating of 100\%.
\subsection{Description of PCA Components}
In order to describe the cluster, we  first describe each PCA component by the features based on loading score within the threshold value.

\subsubsection{PCA Component 1 -- Preference of Non-workstation devices for both Completion and Management}

This PCA component as shown in  Table ~\ref{tab:pca_component_1_features} has a bucket of features focusing on better support for managing HITs on non-workstation devices including tablet phone, watch and speaker with  absolute value spanning from 0.19-0.21.
The second group, with absolute values ranging from 0.10 to 0.18, shows less interest towards not using non-workstation devices for completing HITs in future and especilaly desire to use phone and tablet for completing survey type HITs.
It highlights the versatility of using 'magic wand' for managing HIT functionalities across any device, such as script availability and filtering HITs on any device especially emphasizing tablet and phone.
This component also reflects that  the mentioned magic wand particularly would like to enhance the UX on any device type, pushing the boundaries of conventional workstation-based crowdwork.

Overall, this PCA component underscores the potential for improving HIT management and completion on non-workstation devices and highlights the innovative possibilities of supporting scripts and making UX bettter  in enhancing usability across multiple devices.

\subsubsection{PCA Component 2 -- Selective Device Utilization for HIT Management and Completion}

This PCA component as shown in Table ~\ref{tab:pca_component_2_features} has two buckets.
The first bucket, with feature values ranging from 0.18 to 0.20, indicates a positive rating  for phones and tablets for both managing  HITs. However there is a less enthusiam in using magic wand for supporting script on phones for managing HITs using phone.
The second bucket with absolute value spanning from 0.10-0.17 indicates a positive rating for tablet and phone for completing HITs.
In addition to this, this bucket also reflect the reluctance to use watch, tablet and speaker for completing HITs.
Additionally, this bucket reveals lesser enthusiasm for employing the 'magic wand'  across all devices for supporting scripts to manage HITs epecially on tablet and also shows less enthusiam for better support in managing HITs using tablet.

Overall, this PCA component highlights a general preference for using phones and tablets for management and completion tasks, while indicating less interest in  supporting features like scripts or tools  for managing HITs across any devices, particularly on phone and tablets.

\begin{table*}
    \centering
    \begingroup 
        \fontsize{9}{10}\selectfont 
	\begin{tabular}{@{}l| c| c| c| c@{}}
            \textbf{Features} & \textbf{Statistic} & \textbf{\makecell{$p$\\value}} & \textbf{\makecell{Effect\\ size}} & \textbf{\makecell{Corrected\\ effect \\size}} \\ \hline
            
            \makecell{Desired\_support\_Management\_Tablet }& $\chi^2(12, 148)=89.1$ & $<0.001$ & 0.55 & 0.51 \\ \hline
            \makecell{Desired\_support\_Management\_Phone} & $\chi^2(12, 148)=77.6$ & $<0.001$ & 0.51 & 0.47 \\ \hline
            \makecell{Rate\_Tablet\_Current\_Management} & $\chi^2(8, 148)=70.9$ & $<0.001$ & 0.49 & 0.46 \\ \hline
            \makecell{Desired\_support\_Management\_Speaker} & $\chi^2(10, 148)=69.0$ & $<0.001$ & 0.48 & 0.45 \\  \hline
            \makecell{Desired\_support\_Management\_Watch} & $\chi^2(12, 148)=67.0$ & $<0.001$ & 0.48 & 0.43 \\ \hline
            \makecell{Rate\_Tablet\_Current\_Completion} & $\chi^2(8, 148)=59.8$ & $<0.001$ & 0.45 & 0.42 \\ \hline
            \makecell{Rate\_Phone\_Current\_Management} & $\chi^2(8, 148)=59.0$ & $<0.001$ & 0.45 & 0.42 \\ \hline
            \makecell{Rate\_Phone\_Current\_Completion} & $\chi^2(6, 148)=48.9$ & $<0.001$ & 0.41 & 0.38 \\ \hline
            \makecell{Desired\_HIT\_type\_Survey\_Tablet} & $\chi^2(2, 148)=20.0$ & $0.007$ & 0.37 & 0.35 \\ \hline
            \makecell{Rate\_Watch\_Current\_Management} & $\chi^2(6, 148)=35.1$ & $<0.001$ & 0.34 & 0.32 \\ \hline
            \makecell{Magic\_wand\_HIT\_Management\_script\_Phone} & $\chi^2(2, 148)=17.1$ & $0.028$ & 0.34 & 0.32 \\ \hline	
	\end{tabular}
    \endgroup

    \caption{All the significant features along with their p-value within threshold effect size of more than 0.30 across all three clusters of crowdworkers}
    \label{tab:significant_features}
\end{table*}



\begin{figure}[t]
    \centering
    \includegraphics[width=1.08\linewidth]{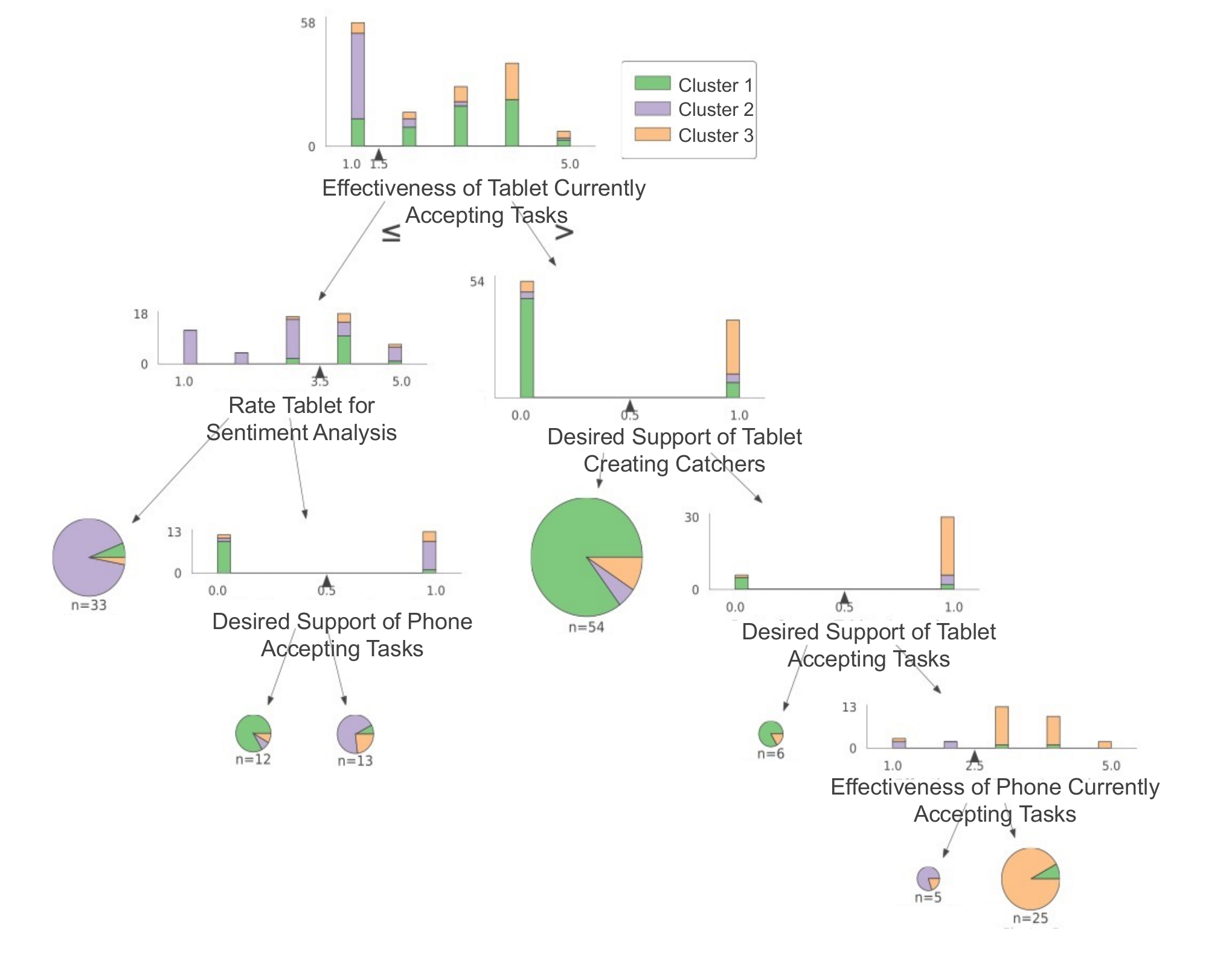}
    \caption{Decision Tree displaying all features that differentiate crowdworker clusters.}
    \label{fig:decision_tree_turk}
\end{figure}

\begin{figure}[t]
	\centering
     \includegraphics[width=\linewidth]{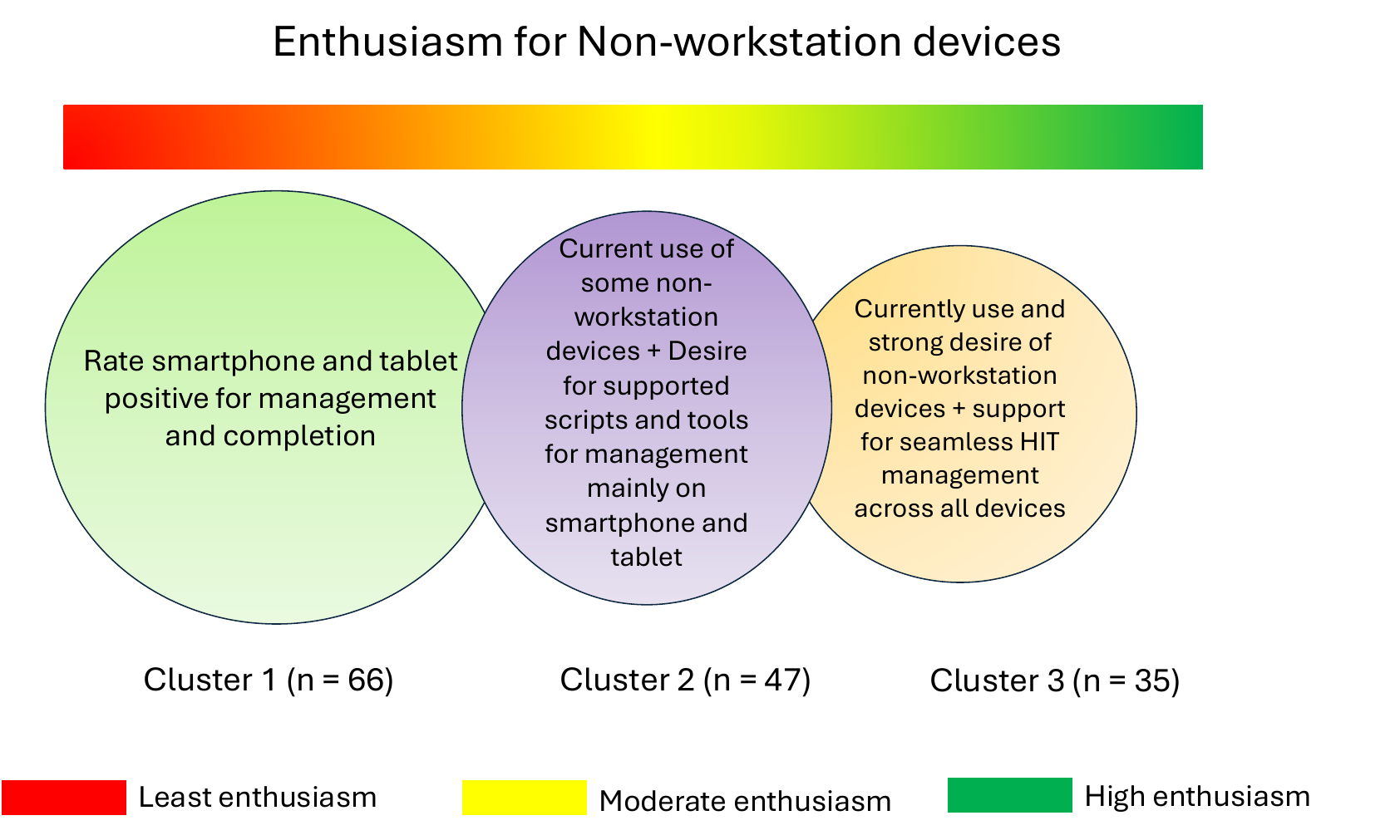}
	\caption{Visualization of crowdworker clusters based on their  enthusiasm for non-workstation devices for task completion and management}
	\label{fig:three_clusters}
\end{figure}

\subsection{Description of Clusters of Crowdworkers}
Here, we describe the three clusters of workers based on their current work practices and preference of using devices on different stages of work as the following by correlating each PCA component based on their mean and median value for PCA based analysis, statistical difference and decision tree.



\subsubsection{Cluster 1 -- Traditionalists with Limited Use and Openness of Smartphone and Tablet}

This cluster of 66 workers demonstrates a nuanced interplay of device preferences for completing and managing Human Intelligence Tasks (HITs), as captured by PCA components 1 and 2 as shown in Table ~\ref{tab:pca_component_mean_median_along_with_cluster}.

PCA component 1, with its higher negative magnitude (mean = -0.86, median = -1.01), indicates a clear low enthusiasm for better supporting tablets, phones, watches, and speakers for management. 
For smartphones and tablets, this group of workers generally shows low enthusiasm for better supporting management tasks, particularly for accepting HITs on phones and creating catchers on tablets, as illustrated in Figure ~\ref{fig:decision_tree_turk}.
Additionally, these workers have no desire to complete HITs using any of these non-workstation devices. This sentiment is echoed in the following participant quotes:
\begin{quote}
	\textit{``I honestly prefer to do all of my work on a laptop. I only use a smartphone if it is absolutely required for the HIT. So I don't really want to see any further support for it.''}(P5)
\end{quote}
\begin{quote}
	\textit{``I don't care for working on my tablet.''}(P74)
\end{quote}
\begin{quote}
	\textit{``I would not like to see any types of HITs better supported on the smart speaker''}(P71)
\end{quote}
\begin{quote}
	\textit{``I do not believe it would be possible to complete HITs on a smartwatch''}(P100)
\end{quote}
PCA component 2, which shows a positive direction (mean = 0.70, median = 0.74), reveals a favorable perception of phones and tablets for completing and managing related work. 
The majority of this group rates tablets and phones highly, between 3 and 4 on a 5-point scale where 5 refers to most suitable, for managing and completing HITs. Despite these high ratings, they currently mostly use smartphone for those HITs which requires to use phone and tablets for completing survey HITs and only if the task requires to use tablet. Majority of the workers in this group do not use watches and speakers for completing HITs currently: 
\begin{quote}
	\textit{``Sometimes I will accept a HIT that needs to be completed on a smartphone \dots''}(P5)
\end{quote}
\begin{quote}
	\textit{``I do alot of surveys on my tablet.''}(P35)
\end{quote}

Despite some positive ratings for phones and tablets in specific scenarios, this group of workers  prefer to avoid all non-workstation devices for completion and have less enthusiam in any better support for managing HITs on the non-workstation devices. 
This indicates a primary reliance on traditional work setting like workstation,  with limited current use and openness of mobile devices including smartphone and tablet. 

\subsubsection{Cluster 2 --Traditionalist with Some Current use of Non-Workstation Devices and a Desire for Management Support mainly on Smartphone and Tablet}

This cluster of 47 workers, characterized by their distinct device usage patterns for tasks, exhibits specific preferences as detailed in PCA components 1 and 2 as shown in Table ~\ref{tab:pca_component_mean_median_along_with_cluster}.
PCA component 2, the more dominant with a higher magnitude and a negative direction, (mean = -1.32, median = -1.23) indicates a low rating  for phone tablet for completing and managing HITs as well as watch for managing HITs .
The workers in this group rate tablets and phones between 1 and 2 on a 5-point scale for management and between 2-3 for completion of HITs where 1 refers to least suitable and 5 refers to most suitable.

They have some openness for using non-workstation devices currently for completion as they are currently using tablet, watch and speaker for completion and using phone for completing phone specific HITs :

\begin{quote}
	\textit{``I only complete HITs which say they require a smartphone to complete \dots''}(P85)
\end{quote}

\begin{quote}
	\textit{``Generally easy ones on [tablet] if I'm away from my computer and I'm relaxing or watching TV.''}(P10)
\end{quote}

\begin{quote}
	\textit{``I often use my smart speaker to get extra information to complete HITs.''}(P32)
\end{quote}

\begin{quote}
	\textit{``I use [ my watch] to bring in information to help with HIT's when they take up my whole screen.''}(P32)
\end{quote}

PCA2 also reflects that this group wants enhanced support features like scripts or tools on all devices, including workstations, phones, tablets, speakers, and watches, with a particular emphasis on phone and tablet. 
They also express a desire for better support in managing HITs on tablets:


\begin{quote}
	\textit{``I would love to have an all in one script that allows me to accomplish all of my tasks unlike now where I need to have multiple scripts and screens open in order to accomplish my work [ on workstation].''}(P6)
\end{quote}

\begin{quote}
	\textit{``Similar to a smartphone, I'd like to see more scripts and tools that help catch work on tablet.''}(P84)
\end{quote}

\begin{quote}
	\textit{``Make  smart speaker easier to accept HITs and work on them or even find good work.''}(P124)
\end{quote}

\begin{quote}
	\textit{``I would make a HIT catcher for portables. To automatically catch HITs.''}(P99)
\end{quote}

PCA component 1, though showing a lesser magnitude and negative direction (mean = -0.12, median = -0.25), corroborates reluctance of non-workstation devices for both completing and managing HITs , which implies the continued use of traditional workstations, highlighting a conventional approach to managing and completing HITs. 

This cluster prefers traditional workstations for both managing and completing HITs highly. In addition to  that they are currently using non-workstation devices like phones, tablets, watches, and speakers for specific tasks. The dominant sentiment is a desire for enhanced support features such as scripts and tools, particularly on phones and tablets, to facilitate HIT management better.  This suggests a hybrid approach where traditional work setups are preferred, but there is a recognition of the potential benefits of using non-workstation devices for management with appropriate support tools.

\subsubsection{Cluster 3 -- Non-workstation Enthusiasts with Seamless HIT Management}
			This cluster of 35 workers, characterized by their distinct device usage patterns for tasks, exhibits specific preferences as detailed in PCA components 1 and 2 as shown in Table ~\ref{tab:pca_component_mean_median_along_with_cluster}.
			
			PCA component 1, with the highest absolute magnitude and a positive direction (mean = 1.79, median = 1.74), indicates a strong preference for better support on all four studied  non-workstation devices, especially phones and tablets, for managing HITs.
			They also mention their high desire for incorporating non-workstation devices for completing HITs. This preference is highlighted by the following participant quotes:
			
			\begin{quote}
				\textit{``I really enjoy HITs where I have to speak phrases into my microphone, but those HITs are often not compatible for a smartphone, even though it would be much easier to do those types of HITs on a smartphone.''}(P12)
			\end{quote}
			\begin{quote}
				\textit{``All HITs should be better suited for a tablet \dots ''}(P33)
			\end{quote}
			\begin{quote}
				\textit{``Transcription HITs would be well-suited for a smart speaker \dots. ''}(P21)
			\end{quote}
			\begin{quote}
				\textit{``I'd love to see more testing HITs on my smartwatch, like ones related to fitness or things like that. I love testing new apps, and I'd love to test ones for HITs on my smartwatch.''}(P12)
			\end{quote}
			
			These workers also want support for scripts for managing HITs  as well as filtering HITs on any device, especially on phones and tablets, and seek better HIT UX compatibility across all studied devices. This sentiment is further reflected in the following participant quotes:
			\begin{quote}
				\textit{``I would use my magic wand to create a more intuitive interface for completing HITs on a tablet \dots''}(P21)
			\end{quote}
			\begin{quote}
				\textit{``I would add HIT catchers and scripts to run on the smartphone platform \dots ''}(P50)
			\end{quote}
			\begin{quote}
				\textit{``Having a version of the web version optimized for a tablet that allows for you to run scripts and tasks.''}(P56)
			\end{quote}
			\begin{quote}
				\textit{``I would setup tools for requesters to make HITs that can be completely answered using a smart speaker \dots ''}(P30)
			\end{quote}
			
			\begin{quote}
				\textit{``On the smartwatch, I would allow apps to find and catch HITs. Workers may be away from their workstations but can carry the watch with them, allowing them to catch HITs while they are away.''}(P62)
			\end{quote}
			
			PCA component 2, though possessing less magnitude (mean = 0.46, median = 0.66) compared to PCA component 1, still contributes valuable insights.
			It shows that workers rate phones and tablets for both managing and completing tasks between 3-4 , where rate watches and speakers between 1-2 in a 5 point scale 5 represents high suitability and 1 represents low suitability. 
           This cluster shows a strong preference for using non-workstation devices like phones and tablets for both completing and managing HITs.
            There is a clear need for greater flexibility and functionality on mobile and other non-workstation devices, emphasizing the importance of versatile and optimized interfaces.\\
			In summary, the clusters differ in their acceptance and enthusiasm towards non-workstation devices for completion and management, with Cluster 1 showing the least enthusiasm, Cluster 2 being moderately open, currently use non-workstation devices along with with a desired support on smartphones and tablets for management, and Cluster 3 being the most receptive and enthusiastic about fully integrating non-workstation devices for managing and completing HITs (see Figure ~\ref{fig:three_clusters}).

\section{Discussion}


\subsection{Design Guidelines Based on Clusters of Crowdworkers’ Work Practices and Preferences}
Here we present design guidelines derived from clustering analyses of crowdworkers, highlighting distinct patterns in device usage for task management and task completion. By understanding these clusters, we offer targeted design recommendations that enhance platform usability, improve task performance, and cater to the unique needs of different crowdworker segments.

\subsubsection{Optimize Desktop Interfaces}
Our findings reveal that the largest cluster of crowdworkers shows a strong preference for traditional work settings, focusing primarily on desktop usage. 

Therefore, it is crucial to prioritize robust functionality and intuitive navigation on desktop platforms, ensuring minimal latency and high performance for desktop-based tasks to maintain user satisfaction and productivity~\cite{santos2021towards, sefati2023ultra}. 
For example, consider a crowdworker engaged in complex data analysis tasks. They rely on their desktop for its processing power and large screen to handle multiple datasets simultaneously. To support this, the platform should provide features like advanced data visualization tools, seamless multitasking capabilities, and a highly responsive interface. Minimal latency in loading data and executing commands ensures the worker can maintain a smooth workflow, thereby enhancing productivity and overall satisfaction. Prioritizing these aspects will help create an efficient and user-friendly desktop experience for crowdworkers.

\subsubsection{Develop Hybrid Systems}
The two largest cluster of crowdworkers relies mostly on traditional workstations but also shows currently use and shows some openness towards using non-workstation devices, especially phones and tablets, for crowdwork.

Therefore, it is essential to create systems that allow seamless transitions between desktops and non-workstation devices, ensuring tasks initiated on desktops can be easily continued on non-workstation devices like smartphones and tablets, and vice versa. This will help support the cognitive processes of workers distributed across different devices~\cite{streitz2001roomware,glodek2015fusion}.
For example, a crowdworker might begin their day by conducting detailed data entry on a desktop. Later, they may need to leave their workstation but want to continue working. With a seamless transition system in place, they can switch to their smartphone to review and edit the data entry while on the go. Features like cloud-based synchronization and responsive design ensure that all progress is saved and the user interface remains consistent across devices. This capability allows the worker to maintain productivity and cognitive flow, regardless of the device they are using, enhancing overall efficiency and satisfaction.

\subsubsection{Enhance Mobile-First Features and Implement Advanced Mobile Capabilities}

Our findings show that the non-workstation enthusiasts group of workers is currently utilizing and seeking enhanced support for non-workstation devices in both managing and completing HITs, aiming for greater efficiency and adaptability in their workflow, mostly on smartphones and tablets. 

Therefore, it is important to ensure that mobile interfaces are intuitive and easy to navigate by understanding the usability requirements of mobile devices~\cite{dutta2022mobilizing,terrenghi2005usability,iwata2010towards} to enhance user experience and usability on the platforms.
For example, a crowdworker might use their smartphone to quickly browse and accept new tasks during a commute. An intuitive mobile interface with large, touch-friendly buttons and streamlined navigation helps them find and accept tasks with minimal effort. Additionally, implementing task management tools that enhance mobile device functionality, such as voice-activated commands, can further improve efficiency. Imagine a worker using voice commands to start a new task or check their task list while cooking or sitting idle.
Moreover, ensuring mobile devices can handle complex tasks with multitasking support and enhanced processing power is crucial. For instance, a tablet might be used to simultaneously handle data entry and video reviews, with the platform dynamically adapting to the user's needs to provide a seamless experience. These innovations help workers maintain high productivity and adaptability in their workflow, leveraging the strengths of their preferred devices~\cite{kwon2022machine}.

\subsubsection{Incorporate Tools for a Smooth Workflow Across Devices}
Our findings reflect an inclination for enhanced management on mobile devices, especially phones and tablets, from the second-largest cluster of workers. Additionally, cluster 3 shows a preference for enhanced management across all devices, from workstations to non-workstation devices.\\
Therefore, it is important to provide multitasking management tools in crowdsourcing platforms that allow workers to manage and complete tasks efficiently on any device~\cite{mcfarlane2002scope,lv2022deep}. 
For instance, a crowdworker might start their day on a workstation. As they move through their day, they can switch to their smartphone to quickly accept new assignments or communicate with requesters using voice commands while waiting for an appointment. Later, they might use a tablet during a commute to review and organize tasks, utilizing a mobile-optimized tool that supports drag-and-drop functionality.
Introducing real-time collaboration tools optimized for specific device use, such as instant messaging and shared document editing on smartphones, and developing streamlined task-switching capabilities, can further facilitate smooth workflow transitions, supporting efficient and adaptable workflows~\cite{carter2021there,ekandjo2024human}. 

\subsection{Understanding Crowdworkers Through the Lens of Traditional and Established Worker Groups}
Our investigation uncovers interrelated patterns of device usage across work stages in crowdwork, including task completion and management. Comparing these practices with traditional 9-to-5 employees enriches our comprehension of contemporary work dynamics.\\
The predominant group of traditionalists mirror the conventional 9-to-5 work model, primarily relies on a single device type—echoing the traditional desk-bound employee—yet exhibits some openness towards integrating mobile devices in the workflow. This behavior resonates with the work habits of information workers, who, as noted in prior research~\cite{cecchinato2015working}, typically use workstations for completing tasks but also open towards using mobile devices~\cite{karlson2010mobile,maassen2020future}.

Workers within cluster 2 demonstrate a high degree of strategy, currently using non-workstation devices for some tasks for completion along with that they would like major support for scripts and tools for better management of HITs on all devices including workstation, phone, tablet, speaker and watch with a higher emphasis on phone and tablet. This strategic diversification of device usage not only optimizes their workflow but also aligns with the practices of modern information workers and IT professionals, who adeptly leverage the unique strengths of different technologies to enhance productivity and efficiency~\cite{goth1999mobile,karlson2009working,oulasvirta2007mobile}.

Cluster 3 workers, the non-workstation enthusiasts, exhibit work practices similar to digital nomads. Both groups prefer using mobile devices like smartphones and tablets for managing and completing tasks, valuing flexibility and mobility. This alignment is evident in their reliance on mobile technology for communication, project management, and accessing cloud-based tools~\cite{reichenberger2018digital,nash2021nomadic}. Both benefit from intuitive mobile interfaces, seamless device synchronization, and the ability to work from any location, emphasizing the need for versatile platforms that support these work styles.

\section{Limitations and Future Work}
We acknowledge limitations, including a small sample size of 150 MTurk workers and a focus on Amazon Mechanical Turk, limiting generalizability. Future research should involve larger samples across platforms like Prolific, CrowdFlower, and Upwork. Self-reported data may also introduce recall errors and response bias, requiring cautious interpretation and further validation.

 \section{Conclusion}
We identified three distinct groups of crowdworkers based on their device usage and work stages. Cluster 1 rely on traditional devices for task completion and management,  limited openness to mobile devices.
Cluster 2, similar to modern IT professionals, use some non-workstation devices and seek support for scripts and tools across all devices, especially on smartphones and tablets.
Cluster 3, resembling digital nomads, prefer smartphones and tablets for managing and completing tasks and desire enhanced support across all devices.
Our findings suggest that personalized crowdsourcing platforms, allowing workers to use preferred devices at different work stages, can improve performance, and cater to diverse needs, revolutionizing productivity.
\bibliographystyle{unsrt}  

\bibliography{aaai24}
\end{document}